\documentclass[a4paper]{article}
\usepackage{amsmath,amssymb,amsthm,graphicx,hyperref,authblk}
\usepackage[utf8]{inputenc}
\usepackage[a4paper,margin=2.5cm]{geometry}
\DeclareUrlCommand\email{\urlstyle{rm}}

\title{Study of transverse oscillations in coronal loops excited by flares and eruptions}

\author{Sandra M. Conde C.$^{1}$\footnote{\href{sandra.conde@usp.br}{sandra.conde@usp.br}}, Rekha Jain$^{2}$\footnote{\href{R.Jain@sheffield.ac.uk}{R.Jain@sheffield.ac.uk}}, and Vera Jatenco-Pereira$^{1}$}

\affil{$^1$Instituto de Astronomia, Geofísica e Ciências Atmosféricas, Universidade de S\~ao Paulo,\\
Rua do Mat\~ao 1226, S\~ao Paulo, 05508-090, Brazil.}
\affil{$^2$School of Mathematics and Statistics, University of Sheffield, S3 7RH, UK}

\date{\today}

\begin{document}
\maketitle

\begin{abstract}
	We present measurements of periodicity for transverse loop oscillations during the periods of activity of two remote and separated (both temporally and spatially) flares. The oscillations are observed in the same location more than 100 Mm away from the visible footpoints of the loops. Evidence for several possible excitation sources is presented.  After close examination, we find that the eruptions during the flaring activities play an important role in triggering the oscillations. We investigate periodicities using time-distance, Fast Fourier Transform, and Wavelet techniques. Despite different excitation sources in the vicinity of the loops and the changing nature of amplitudes, the periodicity of multiple oscillations is found to be 4 -- 6 minutes.
\end{abstract}

\section{Introduction} \label{sec:intro}

It has been known for decades, both  theoretically and observationally that Magnetohydrodynamics (MHD) waves govern the dynamics of solar coronal loops and the study of oscillations can yield some properties of the loops (see \cite{REB84,AFSA99,NODRD99}). Coronal loop oscillations with displacement amplitudes of a few Mm are termed {\it large amplitude} oscillations. These oscillations have been found to decay quite rapidly (see \cite{NNV13}). On the other hand, oscillations that are  small amplitudes and persist for long time without any significant decay are called {\it small amplitude} oscillations \cite{NNV13,ANN15}.\\

Although the observational evidence of large and small amplitude oscillations of varying periodicities have been mounting, understanding of the exact excitation process or even the actual source that excites these flare-induced large amplitude decaying oscillations and the long duration small-amplitude oscillations requires more scrutiny. There have been theoretical efforts to explain the rapid decay of large amplitude oscillations and several mechanisms, such as phase mixing \cite{OA02}, resonant absorption \cite{GAA02,RR02,HJ18}, the wake patterns of a travelling disturbance (see e.g. \cite{TOB05}), and interference phenomena in multi-dimensional wave-cavity \cite{HJ14} have been put forward. Out of these mechanisms, resonant absorption has attracted a lot of attention for the cause of wave damping as coronal plasma is non-uniform and therefore, it is believed that resonant absorption, a conversion of one type of wave into another, is inevitable.\\

\cite{TGV10} investigated an analytical model about the resonant absorption as the damping mechanism. It has also been suggested by \cite{HJ14} that a fast wave propagating from a flaring site can perturb loops in an arcade and set up resonant oscillations in them due to constructive interference effect. Such a scenario does not require any local dissipation because once the fast waves have passed by, the amplitude automatically dies out. Of course, there are loops of varying lengths and shapes in an arcade, and loop oscillations in an arcade can have vertical and horizontal oscillations coupled (see for example, \cite{HJ15,TJ17}). Similarly, for low-amplitude ``decay-less'' oscillations, \cite{HJ14} also demonstrated that the resonant oscillations set up by continuous stochastic sources in the background plasma can make loop structures to oscillate without decay for a long time. However, highly inhomogeneous plasma in the solar corona and the limited spatial resolution of the current instruments make it difficult to see clearly the geometry of the magnetic active region and verify or discard different theories.\\  

Most observational reports of loop oscillations have been concerned with oscillations induced by a single driver (i.e. one visible flare or an eruption near the loop) but in recent times, some interest has been shown in the influence of consecutive flares on oscillations \cite{AJH19,ZDX20} but in general, the literature is limited on this issue. In particular, there is no study, to our knowledge, that has reported oscillations in the same location in an active region due to multiple excitation sources.\\

In the present study, we provide evidence of consecutive oscillations that are observed during the two big successive flaring activity along with other eruptions. Although the flares and eruptions are remote from one another, measurement of oscillations is on the same location. This text is distributed as follows. In Section \ref{sec:obs} we present the observational data and characterize the impulsive events occurring in the active region. In Section \ref{sec:res} we describe in detail the time-distance, Fast Fourier Transform, and wavelet techniques used to detect oscillations in the loops. Also, we discuss the interaction between eruptions, flares, and oscillations found in the loop structure. Finally, we present the conclusions of our study in Section \ref{sec:conclu}.

\section{Observation} \label{sec:obs}

We analyse intensity images of the active region AR 1967. This region was located on the southeast limb of the Sun and had a high number of transient events between 27 and 28 January 2014. We chose an observation time from 04:00 to 06:30 UT on 28 January 2014, pointing out six events that occurred in the region as can be seen in Table \ref{t-events}.\\

To carry out this study, we used images from the Atmospheric Imaging Assembly (AIA) \cite{LAB12} instrument onboard the Solar Dynamics Observatory (SDO) \cite{PTC12}. We assembled data, with a spatial resolution of 0.6$^{''}$ and temporal cadence of 12 seconds, in the 131 \r{A}, 171 \r{A}, and 304 \r{A} bandpasses and prepared them by using procedures available in the Solar Software (SSW). In flaring regions the SDO/AIA channels 131 \r{A}, 171 \r{A}, and 304 \r{A} are expected to observe the Fe XXI ($\log T \sim 7.05$ K), Fe IX ($\log T \sim 5.85$ K), and He II ($\log T \sim 4.7$ K) lines respectively \cite{ODMWT10}.  Therefore, we will describe the characteristics of eruptions and flares covering a wide range of temperatures $5\times10^4 - 11.2\times10^6$ K. The detailed Temperature response can be found in \cite{BWTWS11}.\\

\begin{figure}[h!]
\begin{center}	
\includegraphics[width=0.99\textwidth]{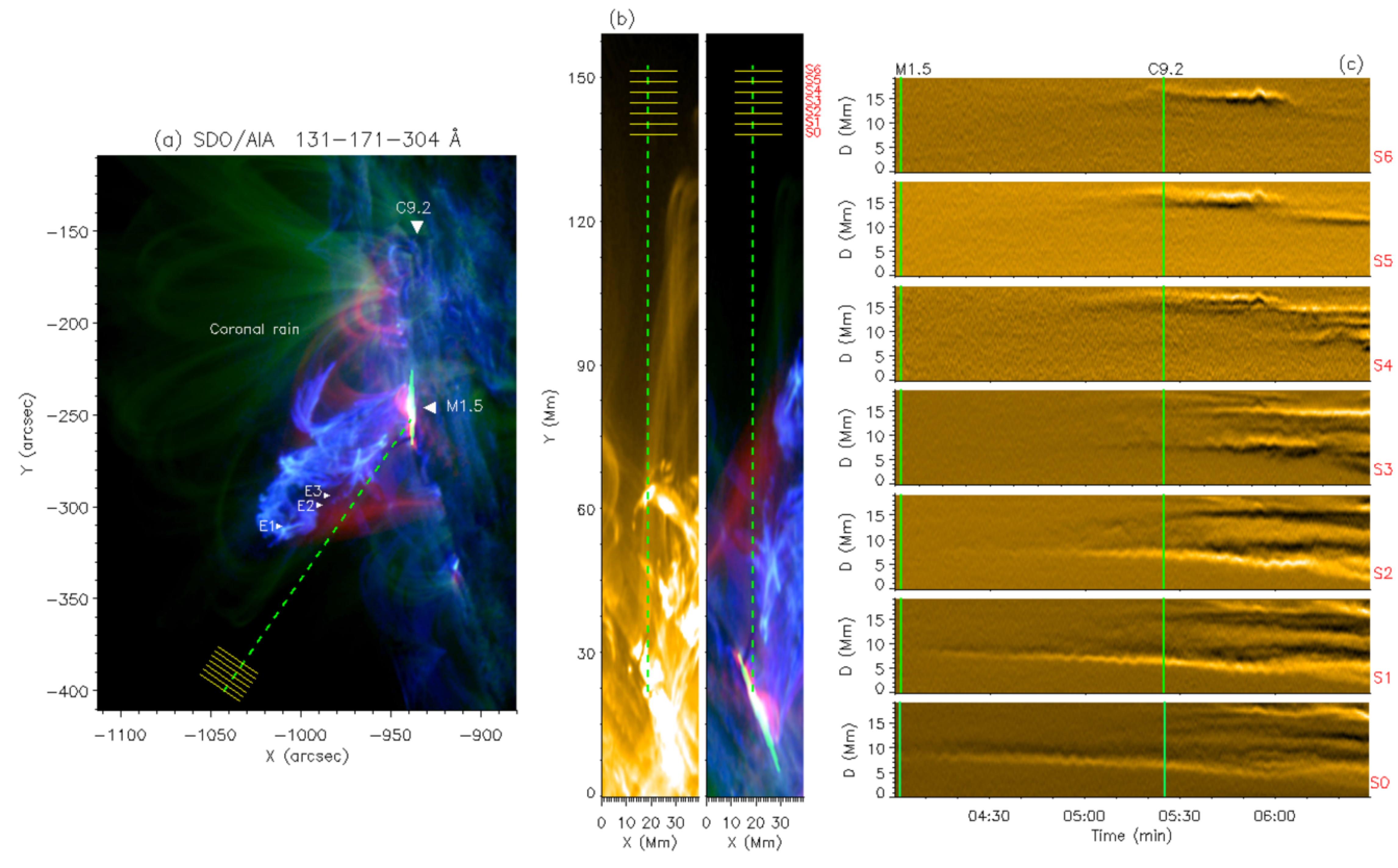}
\caption{Image of the active region AR 1967 obtained by the SDO/AIA instrument on 28 January 2014. (a) SDO/AIA tricolour image in the 131--171--304 \r{A} channels. The dashed green line represents the loops selected for this analysis. The yellow slits (S0 -- S6) on the top of the loops show the region where we found the oscillations. The picture captures the instant the M1.5 flare and eruptions E1 and E2 happened. Also, it is possible to see the location of the C9.2 flare, the eruption E3, and the coronal rain. (b) Image of the loops rotated 146 degrees clockwise in 171 \r{A} (left) and 131--171--304 \r{A} (right). The dashed green line indicates the central axis of the loops. The perpendicular yellow lines represent the slits S0 -- S6. It is possible to see the eruptions reaching the lower part of the loops. (c) Time-distance images of slits S0 -- S6. The green vertical lines represent the beginning of the M1.5 and C9.2 flares. The animation of this figure is available (movgr1967.m4v). It covers the observation period from 04:00  to 06:30 UT on 28 January 2014. The video lasts 18 seconds. \label{fig:mgr1967}}
\end{center}
\end{figure}

The loops selected for this study are clearly visible in 171 \r{A} after 05:30 UT. However, the oscillations were detected from 04:00 UT, when the events 2 -- 4 of Table \ref{t-events} were seen in the vicinity of the loops. In Figure \ref{fig:mgr1967}(a) we present a tricolour image created with the SDO/AIA 304--131--171 \r{A} bandpasses. The green dashed line represents the central axis of the coronal loops selected for the analysis. This image is at the same time the M1.5 flare and the eruptions E1 and E2 were observed. In Figure \ref{fig:mgr1967}(a) we also indicate other events observed in the region: coronal rain, the eruption E3, and the C9.2 flare. The time evolution of the events listed in Table \ref{t-events} can be seen in movgr1967.m4v, which is available on the online version of the journal. Panel (b) of Figure  \ref{fig:mgr1967} shows the region capturing the loops rotated 146 degrees in 171 \r{A} (left) and 304--131--171 \r{A} (right) bandpasses. The yellow slits denoted by S0 -- S6, located around the top of the loops indicate the region of interest (ROI) where we search for oscillations. The intensity variation with time on these slits, as measured in 171 \r{A}, is shown in panel (c).  The full procedure for obtaining such time-distance measurements are described in Subsection \ref{sss:t-d}. \\

\begin{table}
\begin{center}
\caption{Summary of the events in AR 1967 on 28 January 2014.}
\label{t-events}
\begin{tabular}{ccccccc} \hline \hline
No & Event & Time & \multicolumn{3}{c}{Time} & Location ($x,y$)\\
   &       & (UT) & \multicolumn{3}{c}{(UT)} & (arcseconds) \\ \cline{4-6} 
   &       &      & Start & Peak & End       &              \\ \hline
 1 & Coronal rain & 03:48 -- 10:18 & & & & $(-968.3,-235.6)$  \\
 2 & Eruption E1  & 04:00 -- 04:40 & & & & $(-1011.6,-310.8)$ \\
 3 & Eruption E2  & 04:02 -- 04:30 & & & & $(-989.4,-299.1)$  \\
 4 & M1.5 flare   &  & 04:02 & 04:09 & 04:13 & $(-943.8,-229.5)$  \\
 5 & Eruption E3  & 05:19 -- 05:59 & & & & $(-985.2,-294.0)$  \\
 6 & C9.2 flare   &  & 05:25 & 05:29 & 05:31 & $(-935.4,-146.9)$  \\ \hline
\multicolumn{7}{c}{NOTE---Movies and complementary information about the events listed above can}\\
\multicolumn{7}{l}{be found in the websites:}\\
\multicolumn{7}{l}{\url{www.solarmonitor.org}, \url{www.lmsal.com/isolsearch}, and \url{www.helioviewer.org} } 
\end{tabular}
\end{center}
\end{table}

\section{RESULTS AND DISCUSSION} \label{sec:res}

Now that all the main events of AR 1967 have been identified and listed in Table \ref{t-events}, we can examine in detail the oscillations that were prominent in the ROI. This will enable us to understand whether the loop structures in an active region can be excited repeatedly by different events and if so, how does the nature of these oscillations compare.

\subsection{Periodicities of oscillations in a coronal loop structure} \label{sec:osci}

We investigate variations in intensity by taking the time snapshots of the loops seen on January 28, in 171 \r{A} wavelength. Recall that we display the loops in a vertical position by rotating the data cube 146 degrees clockwise (see Figure \ref{fig:mgr1967}(b) left panel). We then traced seven slits (S0 -- S6), indicated by yellow lines, on the top of the loops. They are perpendicular to their central axis (shown by dashed green lines in Figure \ref{fig:mgr1967}(a -- b)) of the loops. Each slit is 19.6 Mm long and 2.18 Mm wide, keeping one pixel ($\approx 0.435$ Mm) of the distance between each one. We also traced other four slits at different heights below the slit S0, but oscillations were not clearly visible in them. So, we focused our attention between S0 and S6 and used different techniques to examine periodicities in these intensity variations. 

\subsubsection{Time-distance diagrams} \label{sss:t-d}

We first create time-distance images to show the time variation in intensity for every slit, increasing the signal-to-noise ratio by smoothing the intensity over the slit width. In order to highlight the oscillatory features, we convolved the image by using the matrix \\

\begin{equation}
	\label{Eq-matrix}	
	\left(
	\begin{array}{rrr}
		1 & 2 & 1 \\
		0 & 0 & 0 \\
		-1 & -2 & -1 
	\end{array}
	\right).
\end{equation}\\

\begin{figure}[ht!]
\begin{center}
\includegraphics[width=0.99\textwidth]{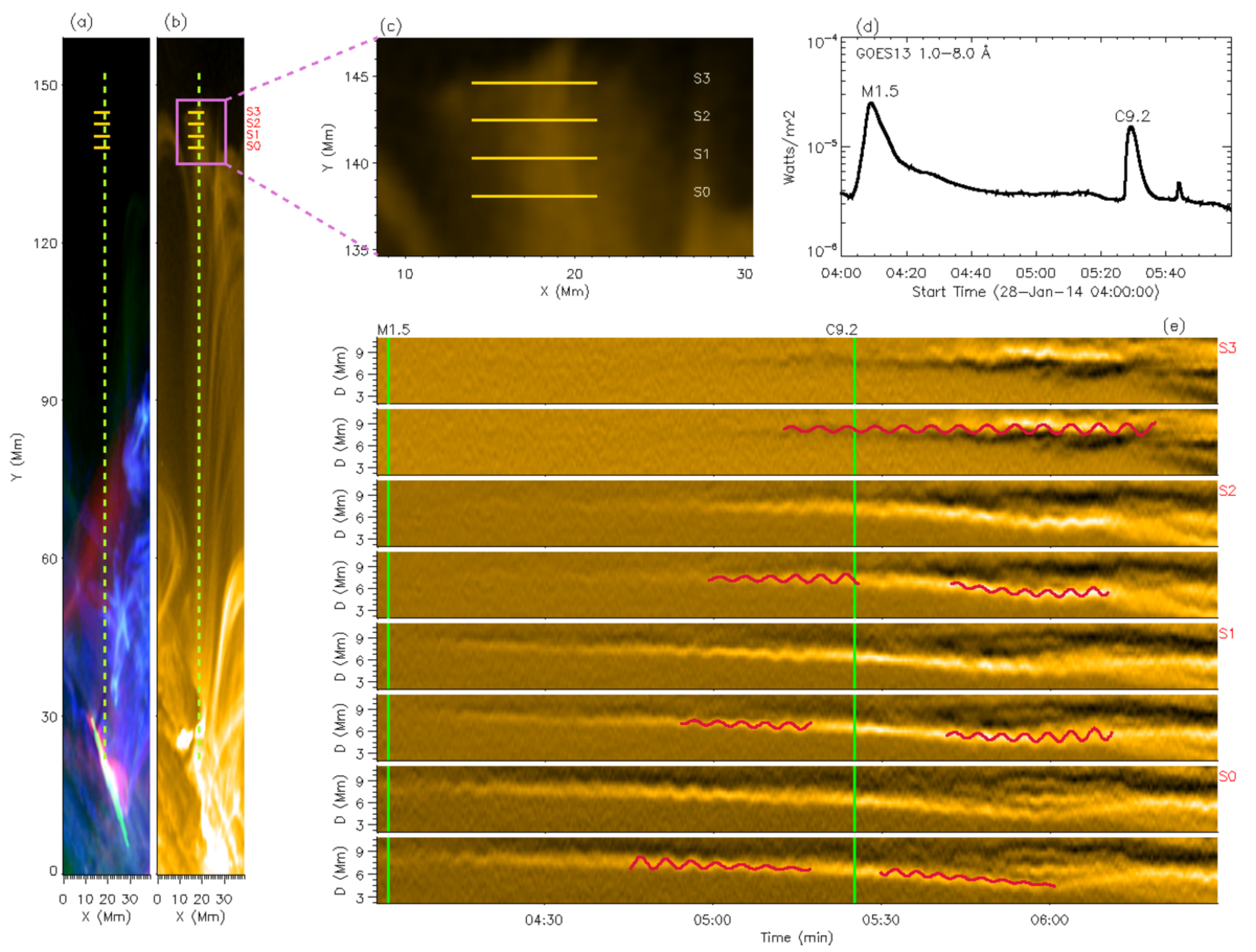}
\caption{Time-distance images along the slits S0 -- S3 traced on the top of the loops. (a) Image of the loops rotated 146 degrees clockwise in the SDO/AIA 131--171--304 \r{A} channels during the M1.5 flare and eruptions E1 and E2. (b) Similar to (a) the image shows the loops in 171 \r{A} after the C9.2 flare starts. The yellow lines represent the part of the slits S0 -- S3 where we found periodicities, which occurred around the central axis of the loops. In panel (c) it is possible to see the region covered by the slits. (d) SXR light curve at 1--8 \r{A} in the time interval between 04:00 and 6:00 UT. The peak flux of the M1.5 and C9.2 flares is visible in the plot. (e) Time-distance images along the slits S0 -- S3. Below each diagram we plotted the same image with the red lines representing the fitted functions for periodicities. The green vertical lines indicate the beginning of the M1.5 and C9.2 flares. \label{fig:GOES}}
\end{center}
\end{figure}

In Figure \ref{fig:mgr1967}(c) oscillatory signals are visible in the lower slits S0 -- S3 soon after 04:02 UT, when the eruptions E1, E2, and M1.5 flare started. Therefore, we now discuss the time-distance in detail, for these slits S0 -- S3. The blow-up view of these four slits, shown in Figure \ref{fig:GOES}(c), illustrates clearly their position on the loop structure. Figure \ref{fig:GOES}(d) shows the soft X-ray flux as a function of time as measured by the Geostationary Operational Environmental Satellite (GOES) \cite{A94goes}. M1.5 flare has a broad distribution of the flux with a peak at about 04:05 UT, whereas the flare C9.2 has a maximum at $\sim$05:30 UT with a narrow distribution of flux. The time-distance plot for each slit is shown in Figure \ref{fig:GOES}(e). At a first glance, it gives the impression of being a continuous oscillating signal which is shifted in time as we move from S0 to S1 to S2 to S3. We do not have any evidence of a wavefront or a blast wave in this dataset during or just prior to these times. Only a coronal mass ejection (CME) was observed by the Large Angle and Spectrometric Coronagraph (LASCO) instrument \cite{BHKK95}, on board the Solar and Heliospheric Observatory (SoHO) spacecraft \cite{DFP95}, on 28 January 2014 between 08:36 and 11:00 UT, but this is outside the time-interval used in our data analysis which is from 04:00 to 06:30 UT.  Even then, it is still possible that there was a coronal propagating disturbance from the lower part of the loop that propagated towards the top of the loop. If this is the case, a propagating disturbance that passes through the slits appears to have a low speed of about 3 km s$^{-1}$, however this speed is likely to be high in reality due to the strong radius of curvature of the loops.\\

A close examination of Figure \ref{fig:GOES}(e) reveals that around 05:30 UT, the oscillations seem to be slightly compressed suggesting that C9.2 flaring activity might have also influenced the same loop system. If we assume that the loop system was disturbed by successive flares (M1.5 and C9.2), we can treat the oscillations to have been excited by the individual flare and extract the oscillation parameters. From the time-distance images, we can extract periods and decay rate by fitting the intensity profile with a Gaussian function. Thus, considering the peak of the Gaussian as the maximum brightness, in a similar way as \cite{CBHNW03}, we fitted the resulting time series with the sinusoidal function $f(t)=A\exp(-t/\tau)\cos(2\pi t/P + \phi)$, where $A$ is the amplitude, $P$ is the period, $\tau$ is the damping time, and $\phi$ is the wave phase. The fitted parameters are shown in Table \ref{t-fit}.\\

It is also interesting to note that there are several visible oscillations close to the main continuous oscillation, especially around 06:00 UT in S0 and S1. Such multiple oscillations on a single slit suggests that the loop structure is inhomogeneous and the slit captures oscillations of several thin nearby loops. Since these oscillations appear to be in phase, it is quite possible that the nearby oscillations are due to skin depth. The near-field response to the main bright oscillating loop depends on the skin depth. Based on a simple model that is commonly used in coronal seismology, \cite{HJ21} have shown that loops of about 160 Mm or more could have a skin depth of 50 Mm or more. Thus, it is possible that here, since the slits S0 -- S3 are within 10 -- 11 Mm and since the main loop is longer than 160 Mm, the loops are coupled and oscillate in harmony.\\

In Figure \ref{fig:GOES}(e) we overlay the fitted oscillations detected on slits S0 -- S3 on the intensity profiles. The horizontal axis is time in minutes and the distance along the slit is denoted by $D$ in Mm on the vertical axis. In S0 the oscillations appear to coincide with the initiation of M1.5 and the eruptions E1 and E2. In Table \ref{t-fit} the fitted parameters show that this is a decaying oscillation with periodicity $P = 4.48 \pm 0.15$ minutes. A few minutes later, we see oscillations of $P = 4.21 \pm 0.07$ minutes in S1, and after 05:00 UT oscillations of $P = 4.5 \pm 0.09$ minutes could also be seen in S2. Contrary to S0, the amplitude of the oscillating threads grows for both S1 and S2. A little before the C9.2 flare begins, some signs of oscillatory signals are visible in S3 but these are really faint. After the C9.2 started, oscillations of periods between $3.84 \pm 0.06$ and $4.99 \pm 0.12$ minutes are seen at $D \approx 5$ Mm, i.e., at the left of the loops' central axis. In S0, the oscillations appear to be decaying, but in S1 and S3 they appear to be increasing in amplitude between 05:45 -- 06:00 UT (see Table \ref{t-fit}).\\

\begin{figure}[ht!]
\begin{center}	
\includegraphics[width=0.99\textwidth]{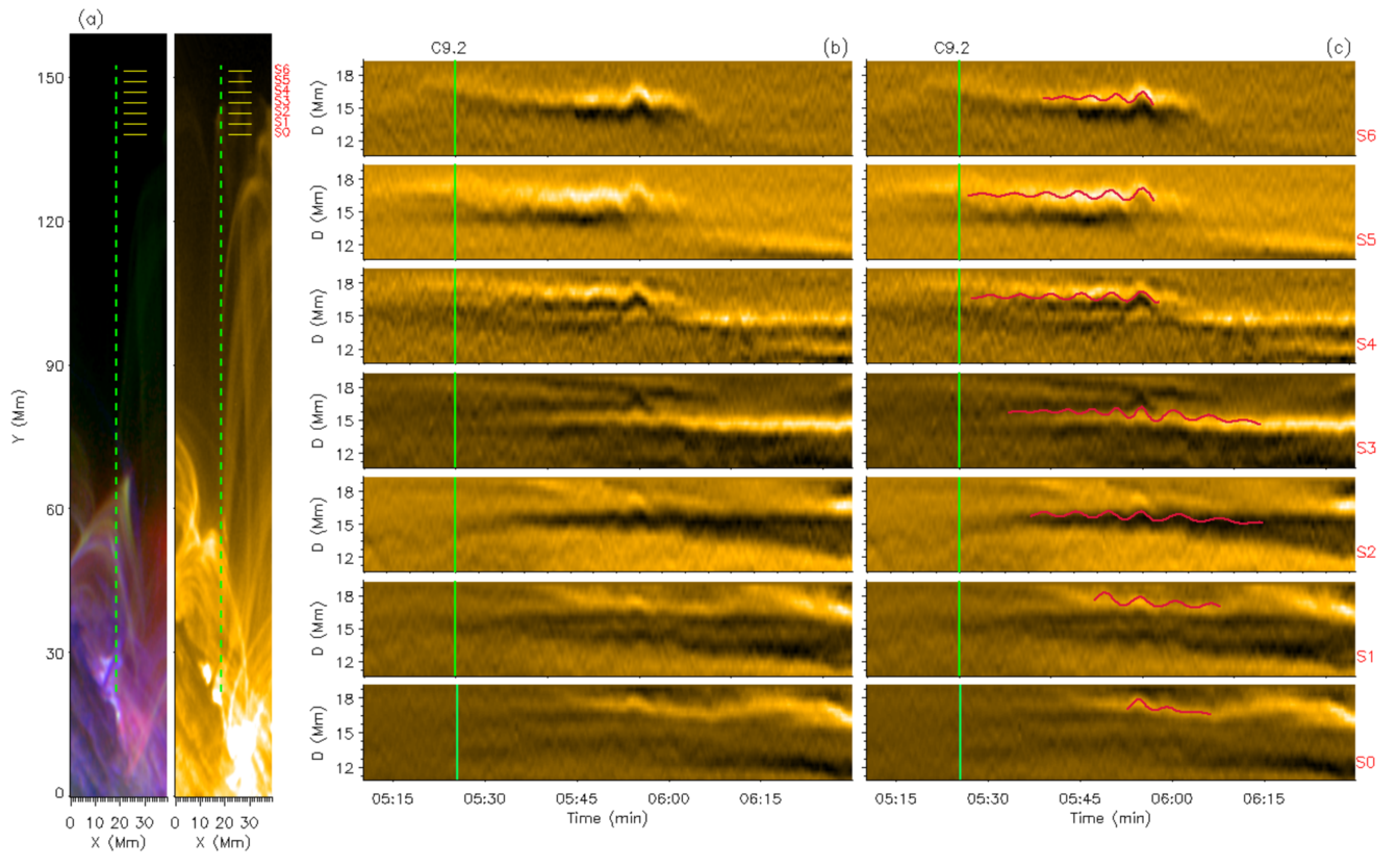}
\caption{Oscillations in the right side of the central axis of the loops after the C9.2 flare started. (a) Images of the loops rotated 146 degrees clockwise in the SDO/AIA 131--171--304 \r{A} (left) and in 171 \r{A} (right) channels. The vertical dashed green line represents the central axis of the loops that we considered. The slit segment analysed in this section is indicated by the yellow horizontal line. (b -- c) Convolved time-distance images along the slits S0 -- S6. The green vertical lines represent the beginning of the C9.2 flare. The red lines overplotted on (c) represent the fitted functions for periodicities. \label{fig:slitc92}}
\end{center}	
\end{figure}

On the other hand, a set of oscillations can be seen clearly in higher slits on the right hand of the loops' central axis around the start of the C9.2 flare at 05:25 UT. In Figure \ref{fig:slitc92}(a) parts of the slits S0 -- S6 are shown by the yellow lines. Also, the eruption E3 is visible in the lower part of the loops. These oscillations are most pronounced at the beginning of the C9.2 flare, near the maximum height of the loops visible in 171 \r{A}, i.e. from S4 to S6. In Figure \ref{fig:slitc92}(b -- c) we show the time-distance diagrams with the time series fitted for slits S0 -- S6. Time-distance images for S0 and S1 show decaying oscillations of $P=4.75 \pm 0.05$ and $P=5.82 \pm 0.07$ minutes. The other slits showed increasing amplitudes with periods between $4.03 \pm 0.08$ and $5.3 \pm 0.08$ minutes (see Table \ref{t-fit}). 
These oscillations with increasing amplitude could be driven by the energy continuously deposited by the C9.2 flare and the eruption E3 that occurred near the footpoint of the loops, which could prevent their rapid damping (see also, \cite{WODS12}). The waveform in Figure \ref{fig:slitc92}(c) around $\sim$05:50 UT appears to be complicated and far from a sinusoidal wave. The shape suggests that the loops may have been excited suddenly and impulsively. The only known event that coincides with this time is the small increase in the X-ray flux, as shown in Figure \ref{fig:GOES}(d).\\

Although the fitted functions give very precise numbers for periodicities (see Table \ref{t-fit}), note that the oscillatory signal is not properly resolved so the periods deduced from the fitting functions are unlikely to be accurate. It is, therefore, sensible to consider periods between 4 and 6 minutes. It appears from Figures \ref{fig:GOES} and \ref{fig:slitc92} that the entire loop system where the slits were placed is influenced by the flares and the associated eruptions. It has been proposed by \cite{HJ15} that in fact, the entire arcade is impacted by flaring activities and to properly understand the connection between oscillations, background plasma, and the flaring activities, we need to look at the characteristics of the large part of the active region. Therefore, we will now examine the full region of interest, as shown in Figure \ref{fig:GOES}(c), instead of just the four slits.

\begin{table}
\begin{center}
\caption{Fitted parameters for periodicities found along the slits S0 -- S6 represented in Figures \ref{fig:GOES} and \ref{fig:slitc92}}
\label{t-fit}
\begin{tabular}{cccccc} \hline \hline
Slit & Class Flare & $A_0$ & $\tau$ & $P$   & $\phi$ \\
     &             & (Mm)  & (min)  & (min) & (deg) \\
 (1) & (2)         & (3)   & (4)    & (5)   & (6) \\ \hline
\multicolumn{6}{c}{Fitted parameters Figure \ref{fig:GOES}}\\ \hline 
S0 & Before C9.2 & 0.93 $\pm$ 0.07 & -17.30 $\pm$ 0.19 & 4.48 $\pm$ 0.15 & -28.39 $\pm$ 6.03 \\
   & After C9.2  & 0.47 $\pm$ 0.02 & -20.24 $\pm$ 0.71 & 3.84 $\pm$ 0.06 &  107.34 $\pm$ 97.17 \\ \hline
S1 & Before C9.2 & 0.29 $\pm$ 0.01 &  46.09 $\pm$ 1.66 & 4.21 $\pm$ 0.07 &  146.89 $\pm$ 80.58 \\
   & After C9.2  & 0.33 $\pm$ 0.04 &  29.13 $\pm$ 0.49 & 4.14 $\pm$ 0.09 &  228.22 $\pm$ 64.17 \\ \hline
S2 & Before C9.2 & 0.26 $\pm$ 0.01 &  26.04 $\pm$ 1.65 & 4.50 $\pm$ 0.09 &  199.18 $\pm$ 61.13 \\
   & After C9.2  & 0.26 $\pm$ 0.04 &  31.46 $\pm$ 0.35 & 4.03 $\pm$ 0.12 & -103.20 $\pm$ 49.16 \\ \hline
S3 & After C9.2  & 0.35 $\pm$ 0.05 &  76.96 $\pm$ 0.14 & 4.99 $\pm$ 0.12 &  -97.17 $\pm$ 49.12 \\ \hline
\multicolumn{6}{c}{Fitted parameters Figure \ref{fig:slitc92}}\\ \hline   
S0 & After C9.2  & 0.66 $\pm$ 0.06 &  -4.78 $\pm$ 0.24 & 4.75 $\pm$ 0.05 & 200.54 $\pm$ 112.46 \\
S1 & After C9.2  & 0.63 $\pm$ 0.03 & -18.08 $\pm$ 0.42 & 5.82 $\pm$ 0.07 & 114.59 $\pm$ 85.42 \\
S2 & After C9.2  & 0.19 $\pm$ 0.01 &  23.57 $\pm$ 1.61 & 5.30 $\pm$ 0.08 & 211.99 $\pm$ 72.74 \\
S3 & After C9.2  & 0.05 $\pm$ 0.02 &   8.75 $\pm$ 0.61 & 4.03 $\pm$ 0.08 & 215.67 $\pm$ 76.58 \\
S4 & After C9.2  & 0.13 $\pm$ 0.04 &  21.38 $\pm$ 0.17 & 4.97 $\pm$ 0.11 & 143.24 $\pm$ 53.88 \\
S5 & After C9.2  & 0.10 $\pm$ 0.05 &  16.23 $\pm$ 0.15 & 5.24 $\pm$ 0.10 & 198.91 $\pm$ 61.23 \\
S6 & After C9.2  & 0.09 $\pm$ 0.04 &   8.89 $\pm$ 0.19 & 4.27 $\pm$ 0.10 &  82.17 $\pm$ 58.65 \\ \hline
\end{tabular}
\end{center}
\end{table}

\subsubsection{Fast Fourier Transform} \label{sss:FFT}

In the previous section, we created the convolved time-distance images for the slits S0 -- S6 for the duration 04:00 -- 06:30 UT.  We detected periodicities between 4 -- 6 minutes by fitting the oscillatory time signals with a sinusoidal function in those convolved images. However, the oscillating signals and the associated periodicities were for selected slits in the ROI. It would be interesting to see whether locations (pixels), other than on the slits in the ROI, show the same periodicities. Recall that we have chosen the top of the loops shown in Figure \ref{fig:GOES}(c) as the ROI.\\

\begin{figure}[ht!]
\begin{center}	
\includegraphics[width=0.9\textwidth]{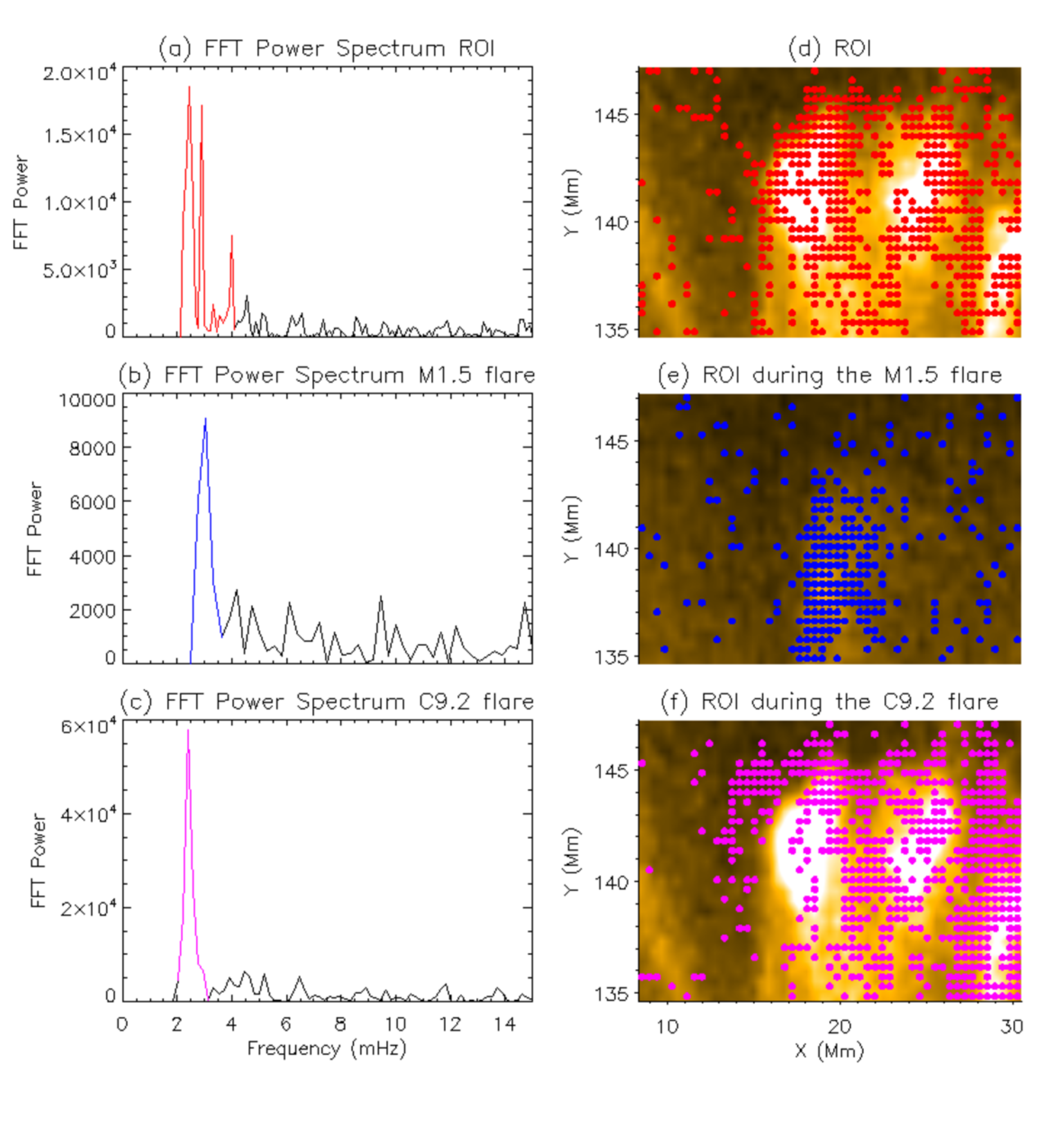}
\caption{FFT analysis of the region at the top of the loops. The left column represents the power spectra found in different time intervals. (a) Between 04:00 and 06:30 UT where the dominant power is in the range of 2.2 -- 4.1 mHz. (b) From 04:00 to 05:00 UT capturing only M1.5 flaring activity. The dominant power is in the band of 2.5 -- 3.6 mHz. (c) Interval of 05:00 - 06:30 UT capturing C9.2 flaring activity. The dominant power is between 2.0 and 3.1 mHz. The right column represents the pixels where the frequencies are dominant. (d -- f) correspond to (a -- c), respectively. \label{fig:FFTgv1}}
\end{center}
\end{figure}

We calculated the Fast Fourier Transform (FFT) of the time series that also includes the peak time of fluxes for flares M1.5 and C9.2. With the aim to understand the dominant frequencies of each flare, we also separate the time series into two different intervals and carry out the FFT for each of these time intervals. In the left column  of Figure \ref{fig:FFTgv1}, we present the FFT power spectrum. The calculated power is the square of the FFT's absolute value. The panel (a) is from 04:00 to 06:30 UT; panel (b) is from 04:00 to 05:00 UT (including only the M1.5 flare) and panel (c) is from 05:00 to 06:30 UT (including only the C9.2 flare).  The red line in panel (a) indicates the frequency band that includes the two prominent peaks in this spectrum. Panel (b) shows peak power at 3 mHz and panel (c) at 2.5 mHz. In the right column, we mark the pixels with the dominant power. The red points in panel (d) show the pixels with dominant power in the frequency range $f=2.2-4.10$ mHz. The blue points in (e), represent the frequency $f=3.05 \pm 0.55$ mHz and the magenta points in (f) correspond to $f=2.58 \pm 0.55$ mHz. \\

As can be seen from Figures \ref{fig:FFTgv1}(e) and (f), the spatial distribution of wave-power is very different during M1.5 flaring activity than C9.2 flaring activity. The dominant power is concentrated in the middle part of the loop structure in the ROI during the M1.5 flare as if the flare excited the entire loop structure coherently in addition to many other faint structures in the active region. During the C9.2 flaring activity, the wave power within the frequency band is distributed quite widely across the loop structure in the ROI. Also, note that the intensity distribution as seen in SDO/AIA 171 \r{A} bandpass is also quite different during the two flaring activities.  Such different intensity and different spatial distributions of power is an indication of inhomogeneous loop structure. \\

It should be noted that although the overall distribution of power can be examined this way, caution is required as FFT analysis can sometimes produce artificial peaks in the power spectrum due to complex nature (e.g. non-sinusoidal signal, especially the type seen in slits S4 -- S6 around 06:00 UT and/or unresolved intensity between pixels etc.) of the time series (see, \cite{AJ21}).  Another issue that requires utmost care is that the start time of C9.2 flare and the decay time of M1.5 flare is not known precisely, there may be some overlap in the two time intervals that we used for the FFT. Thus, FFT 04:00 -- 05:00 UT and 05:00 -- 06:30 UT may introduce cross-contamination of power. Also, a shorter time series of one hour has a poor frequency resolution. These drawbacks warrant data with better temporal and spatial cadence before any other quantitative results can be confidently trusted except for the dominant power around a specific frequency.\\

\begin{figure}[ht!]
\begin{center}	
\includegraphics[width=0.7\textwidth]{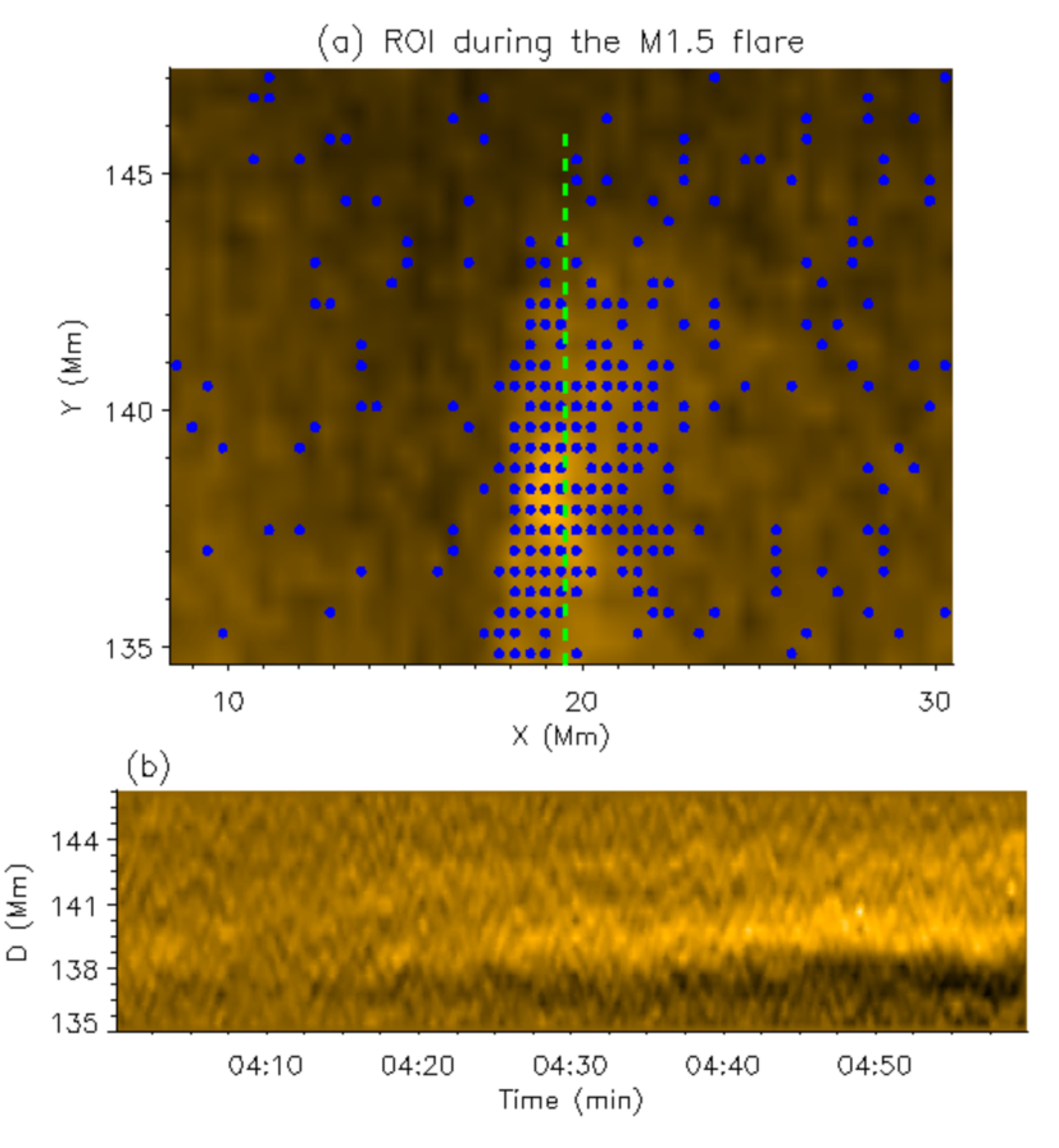}
\caption{FFT analysis and time-distance image in the region during the M1.5 flare. (a) ROI displayed in Figure \ref{fig:FFTgv1}(e). The blue points represent the pixels where the frequency band of  2.5 -- 3.6 mHz is dominant. The dashed green line indicates the slit where we made the time-distance image shown in (b). \label{fig:slitv_lm15}}
\end{center}
\end{figure}

From the spatial distribution of the dominant power as shown in the right column in Figure \ref{fig:FFTgv1}, it is clear that oscillations are not confined to a thin loop or an individual thread. The nearby loops are also either impacted or the excitation source excites more than a single thread. It has been shown by \cite{HJ21} that many coronal loops do not oscillate as an independent entity. The skin depth of an oscillating loop affects the nearby plasmas and field lines. These field lines can oscillate sympathetically in phase in response to the main loop's oscillations due to the skin depth. Since Figure \ref{fig:FFTgv1}(e) suggests that the entire middle region of the ROI has a similar frequency, we analyse the time series on a vertical slit as shown in Figure \ref{fig:slitv_lm15}. The slit is 5 pixels wide and 27 pixels long. Panel (b) shows the resulting time-distance diagram where it is clear that the near-field of the main loop also oscillates in phase.

\subsubsection{Wavelet Transform} \label{sss:WT}

\begin{figure}[ht!]
\begin{center}	
\includegraphics[width=0.8\textwidth]{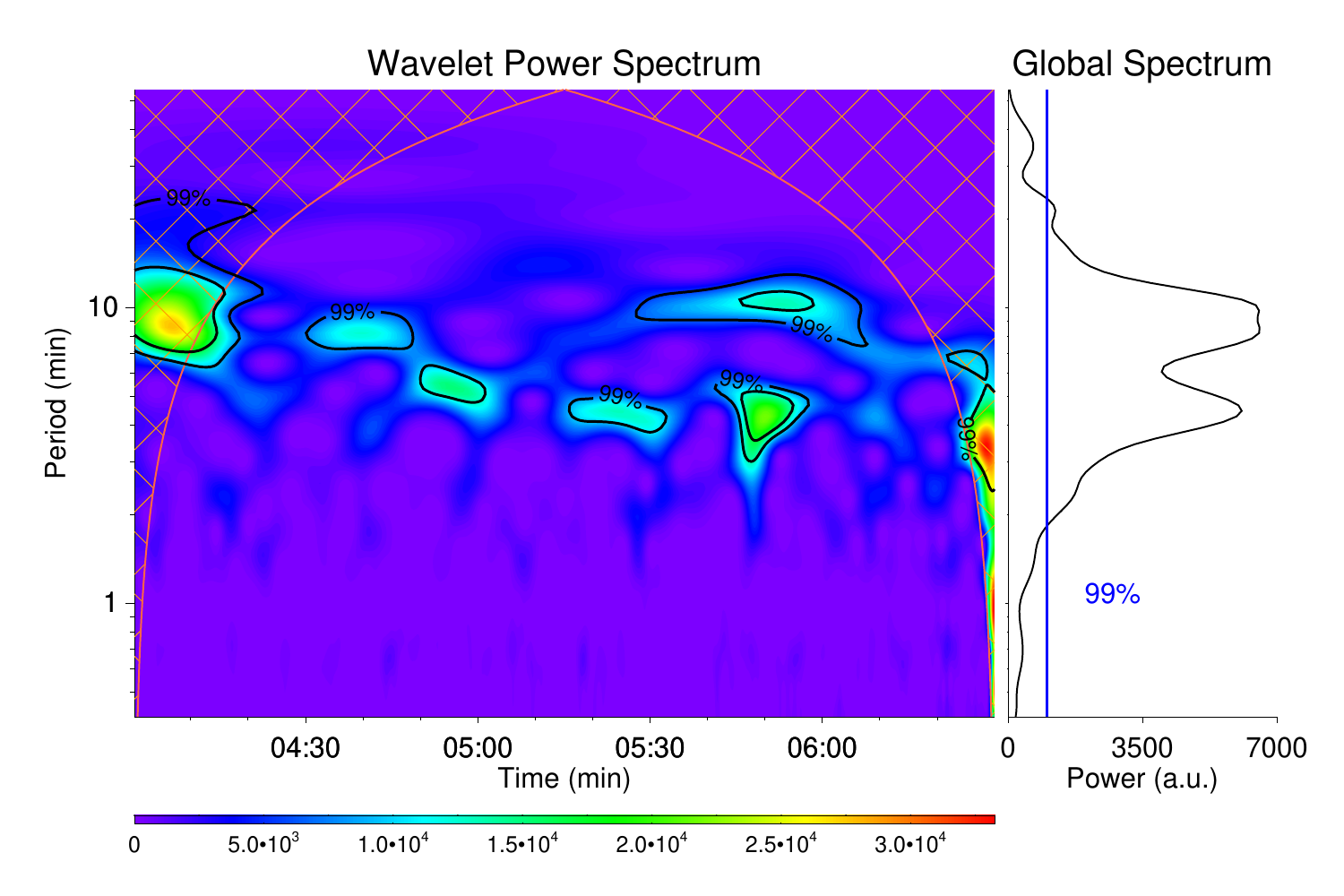}
\caption{Wavelet power spectrum of the time series along the slits S0 -- S6 (see Figure \ref{fig:mgr1967}(c)). The purple curve indicates the limit of the cone of influence (COI). In the same manner, purple cross lines show the part of the spectrum that is inside the COI. The dominant periods of 4.48 and 9.10 minutes are enclosed by the black contours of 99\% confidence level. These periods are also above the 99\% significance level in the global spectrum displayed on the right. \label{fig:wtfd}}
\end{center}
\end{figure}

We now compute the wavelet transform for the time series of the area around the top of the loops, i.e., from S0 to S6. Figure \ref{fig:wtfd} shows the wavelet power spectrum (left panel) for the time series of the area covered for the slits in Figure \ref{fig:mgr1967}(b). The purple line and the dashed region show the area inside the cone of influence (COI). Outside this COI the periodicities above 99\% confidence level are enclosed by black contours, which show dominant periods centred at 4.48 and 9.10 minutes. On the right side, we present the global power spectrum where two peaks indicate the same dominant periods above the blue line, which represents the 99\% significance level. Therefore, comparing the periods from the fitted functions, FFT, and wavelet transform, we can consider $P \approx 4.5$ minutes as the dominant period.

\subsection{Interaction between eruptions, flares, and loops} \label{sec:erfos}

\cite{ZN15} analysed 58 events of oscillating loops that were impulsively excited. They found that out of these, 57 of them were associated with eruptions/ejections in the lower coronal plasma. In the current study also, oscillations are accompanied by eruptions E1, E2 and E3 (see Table \ref{t-events}). However, it is unclear whether the ultimate perturbation to the loops leading to their oscillations is either the displacement of the loop's footpoints from their equilibrium position by these eruptions or the deposition of energy from the eruptions to the loop. If the former, it is not possible to see such a displacement due to limited spatial resolution. If latter, we can estimate the possible maximum speed, $(v_{e})$, of the plasma eruptions that would be necessary to carry energy to perturb the loops. We now discuss this.\\

\begin{figure}[ht!]
\begin{center}	
\includegraphics[width=0.99\textwidth]{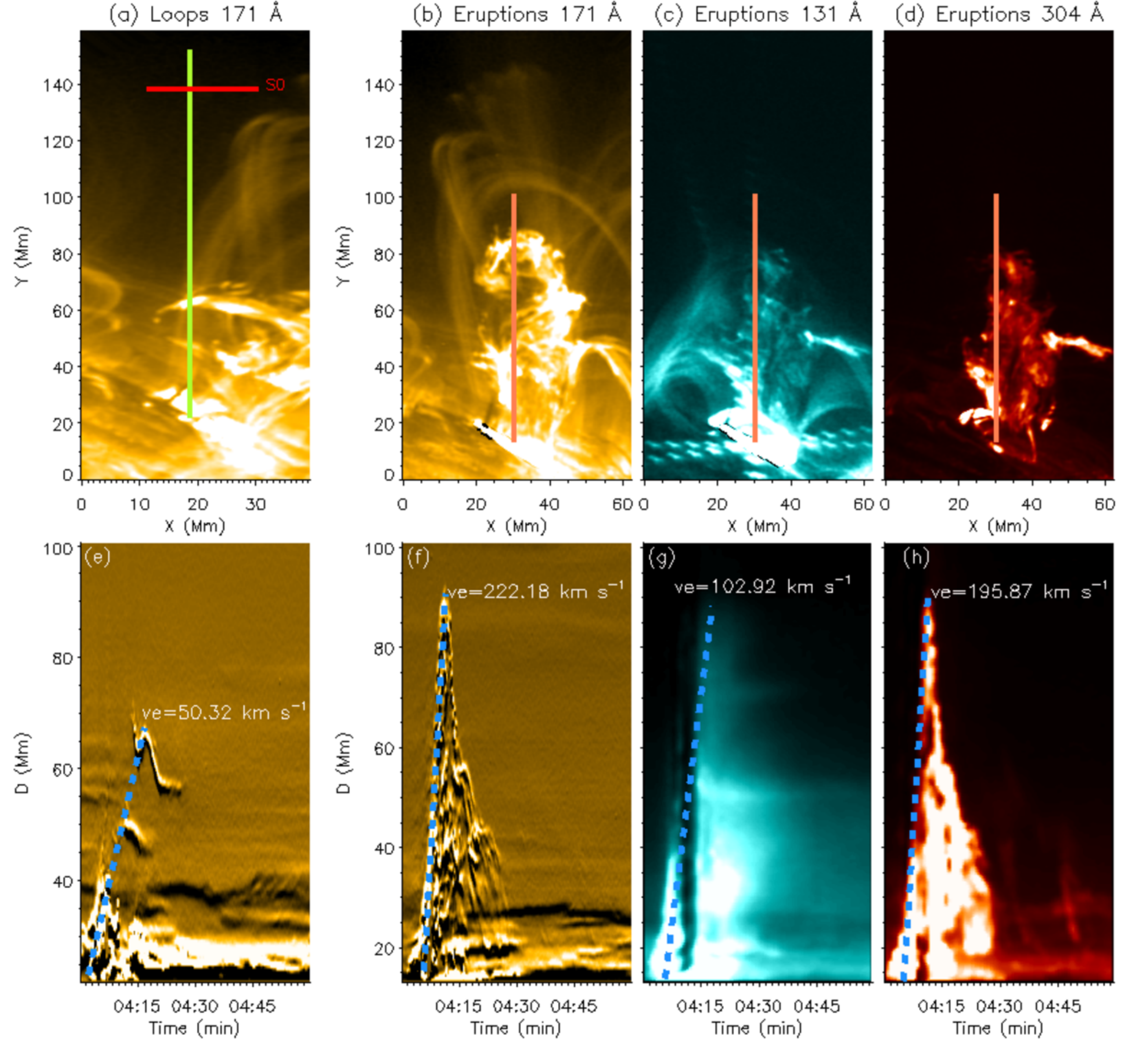}
\caption{Speed of the eruptions E1 and E2 around the loops. (a) Image of the loops in 171 \r{A} at the time of the eruption E2. The central axis of the loops and the slit S0 are represented by the green and red lines, respectively. Images of the eruptions E1 and E2 taken by SDO/AIA in the 171 (b), 131 (c), and 304 \r{A} (d) channels. The orange lines indicate the central axis of the eruptions, which we use to make the time-distance diagrams displayed in the bottom panels. (e -- h) Time-distance images over the slits traced in panels (a -- d). The vertical axis in panel (e) is plotted from 20 to 100 Mm instead 20 -- 150 Mm to show in detail the distance covered by E2. The blue dashed lines indicate the speed of the eruptions E1 and E2 $(v_e)$ along the slits. \label{fig:em15}}
\end{center}
\end{figure}

The events listed in Table \ref{t-events} were also visible in the SDO/AIA 304 \r{A} and 131 \r{A} channels. Since the beginning of the observation time, M1.5 flare and eruptions E1 and E2 may have perturbed the lower part of the loops, which produced periods of $P \approx 4-5$ minutes from slits S0 to S2 (see Figure \ref{fig:GOES}(e)). In Figure \ref{fig:em15} we display time-distance images for slits traced over the central axis of both the loops and eruptions in the SDO/AIA 171 \r{A}, 131 \r{A}, and 304 \r{A} channels. In order to highlight features of the eruption observed in 171 \r{A} we convolved the time-distance images by using the matrix defined in \ref{Eq-matrix}. Figure \ref{fig:em15}(e) displays the time-distance image of a slit 2.18 Mm wide and 130.5 Mm long, along the central axis of the loops (green line in Figure \ref{fig:em15}(a)). We have defined the vertical axis in Figure \ref{fig:em15}(e) from 20 to 100 Mm instead 20 -- 150 Mm, to show in detail the distance travelled by E2. The slopes traced with blue dashed lines indicate the speed of the eruptions $v_{e}=50.32$ km s$^{-1}$. In Figures \ref{fig:em15}(b -- d) the slits are 2.18 Mm wide and 88.43 Mm long. The corresponding time-distance images are present in Figures \ref{fig:em15}(f -- h) where the slopes traced with a dashed blue line indicate the speed reached by the eruptions. They are $v_e=222.18$ km s$^{-1}$, $v_e=102.92$ km s$^{-1}$, and $v_e=195.87$ km s$^{-1}$ in panels (f -- h), respectively. Despite the distance between the central axis of the loops and the eruptions being $\approx 11.75$ Mm, in movgr1967.m4v it is possible to see the M1.5 flare situated in the footpoint of the loops and the eruptions reaching the lower part of them. From the time-distance images over the central axis of the eruptions, we estimated that they reached a height of $\approx 90$ Mm. The distance between the highest point reached by the eruptions and the first slit (S0) was about $50$ Mm.\\

\begin{figure}[ht!]
\begin{center}	
\includegraphics[width=0.99\textwidth]{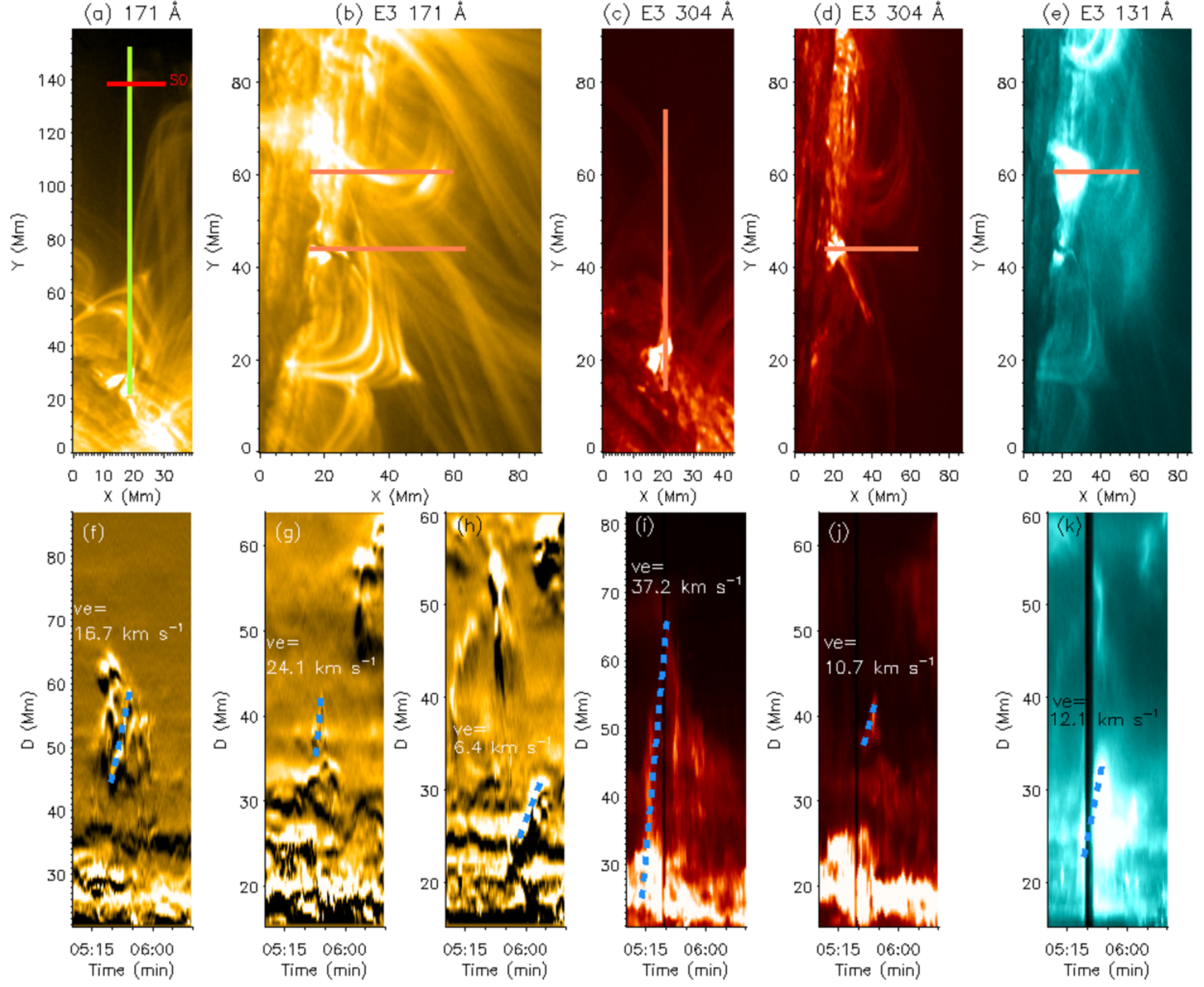}
\caption{Speed of the eruption E3 around the coronal loops. (a) Image of the emerging arcade, in 171 \r{A}, reaching the lower part of the loops a few minutes after eruption E3 began. The central axis of the loops and the slit S0 are represented by the green and red lines, respectively. (b) Flipped image of the loops in the 171 \r{A} wavelength. The orange lines represent the spatial coincidence with the eruption observed in 304 \r{A} (bottom line) and 131 \r{A} (upper line), which are shown in panels (d) and (e), respectively.  (c) Ejected plasma by the eruption E3, observed in a 304 \r{A} rotated 137 degrees clockwise image. The orange line represents the slit traced to make the time-distance map over the central axis of this ejected plasma.  (f -- k) Time-distance diagrams for the panels (a -- e). The blue dashed slopes indicate the speed of the eruption E3 $(v_e)$ along the slits (orange lines). Note the difference in the scales on y-axes, for enhancing the features, in some of the bottom panels. \label{fig:ec92}}
\end{center}
\end{figure}

Just after C9.2 flare started, oscillations with periods of $5-6$ minutes were seen in the right part of the central axis of the loops (see Figure \ref{fig:slitc92}(b)). This flare site was located $\approx 108.9$ Mm away from the loop top, but no wavefront or direct ejection of plasma was visible in the data. Instead, we observed an eruption, E3, that occurred close to the footpoint of the loops. It was then followed by emergence of an arcade in the lower part (see Figure \ref{fig:slitc92}(a)). To see the evolution of the eruption E3, we traced several slits over the images of SDO/AIA in 171 \r{A}, 304 \r{A}, and 131 \r{A} channels, as shown in Figure \ref{fig:ec92}(a -- e). \\

Flare C9.2 started about 6 minutes after eruption E3 (see Table \ref{t-events}). Both events were seen in 171 \r{A}. However, the beginning and the end of E3 were most clearly visible in 304 \r{A} and 131 \r{A}, respectively. Figure \ref{fig:ec92}(a) shows the emerging arcade and E3 at the lower part of the loops. The green and red lines indicate the central axis of the loops and the first slit, S0. These were used for further investigation. We  considered 2.8 Mm wide slit and 130.5 Mm long over the central axis of the loops, to show the time-distance image displayed in the panel (f).  This panel clearly shows the emerging arcade. The slope shown with a blue dashed line indicates that a speed of $v_e \approx 17$ km s$^{-1}$ would be required by the erupting plasma to reach the top of the loops.\\

In Figure \ref{fig:ec92}(b) we displayed the flipped image of the AR 1967 in 171 \r{A}. The orange lines indicate the two triggered points where the eruption E3 was seen. Details of the E3's first part were visible in 304 \r{A} (Figure \ref{fig:ec92}(d)). Thus, we traced a slit 2.8 Mm wide and 48.7 Mm long at the base of the eruption. This was represented by the orange lines in panels (b) and (d). In panels (g) and (j) we show the time-distance diagrams for the slit traced in both 171 \r{A} (panel (b)) and 304 \r{A} (panel (d)). It is possible to see that the plasma reached a height of 42 Mm from the base with speeds $v_e=24.1$ km s$^{-1}$ in 171 \r{A} and $v_e=10.7$ km s$^{-1}$ in 304 \r{A} (see panels (g) and (j), respectively). From this same point, it was ejected plasma in the direction of the loops. Thus, we rotated this image 137 degrees clockwise and traced a slit on the central axis of the ejection (see panel (d)). The time-distance diagram displayed in the panel (i) shows that the plasma covered a distance of $\approx 60$ Mm with a speed of $v_e=37.2$ km s$^{-1}$. The second part of E3 was seen in 171 \r{A} (panel (b)) and 131 \r{A} (panel (e)). We also traced a slit 2.8 Mm wide and 44.78 Mm long, over the central axis of this eruption which was represented by the orange lines in panels (b) and (e). Time-distance images show the evolution of this eruption in panels (h) and (k). The slopes traced estimate the speed of this eruption in 171 \r{A} to be $v_e=6.4$ km s$^{-1}$ (panel (h)) and $v_e=12.1$ km s$^{-1}$ (panel (k)) in 131 \r{A}. \\

Although there was no visible wavefront, even if we speculate that an invisible perturbation reached the top of the loop structure, we can estimate the fastest possible speed. The peak flux for M1.5 and C9.2 was for about 660 and 360 seconds, respectively. Taking into account that the flares were located at $(-943.8'',-229.5'')$ (M1.5) and $(-935.4'',-146.9'')$ (C9.2), and the slit S0 was at $(-1044.76'',-398.38'')$, the speeds can be estimated at $v_{fl}=216.43$ km s$^{-1}$ for M1.5 and $v_{fl}=553.03$ km s$^{-1}$ for C9.2, from the following equation

\begin{equation}
	\label{Eq:vfl}
	v_{fl}=\frac{\sqrt{(x_{osc}-x_{fl})^2+(y_{osc}-y_{fl})^2}}{\Delta t_{fl}},
\end{equation}

\noindent with coordinates of both S0 $(x_{osc},y_{osc})$ and the flare $(x_{fl},y_{fl})$. \\

It has also been suggested that the change in magnetic pressure, leading to implosion \cite{H00} could lead to oscillatory behaviour as the loops loose their equilibrium state during the magnetic reconnection. This possibility cannot be ruled out. However, the observational signatures to support this scenario remain eluded in a resolution-limited dataset as this one. The presence of coronal rain, which is clearly visible in movgr1967.m4v could also be a source for the excitation of oscillations in the loops. However, the change in amplitude with time and their connection to the coronal rain will need to be explored further. 

\subsection{Energy density estimates for the events and the oscillations} \label{sec:enev}

It is useful to investigate further the generation mechanism of oscillations by comparing the energy density produced by the transient events and the oscillations. In a similar way as \cite{SPSB16}, we calculated the kinetic energy density of the eruptions and flares, assuming the linear motion of these events concerning the central axis of the loops by using the equation

\begin{equation}
	E_{ev}=\frac{1}{2} \rho_{ev} v^2,
	\label{E-enev}
\end{equation}

\noindent where $\rho_{ev}$ is the density of the plasma related to the events, i.e., eruptions (E1, E2, and E3) and flares (M1.5 and C9.2), and $v$ is the estimated speed of the plasma (see Figures \ref{fig:em15} and \ref{fig:ec92}).\\

\begin{figure}[ht!]
\begin{center}	
\includegraphics[width=0.7\textwidth]{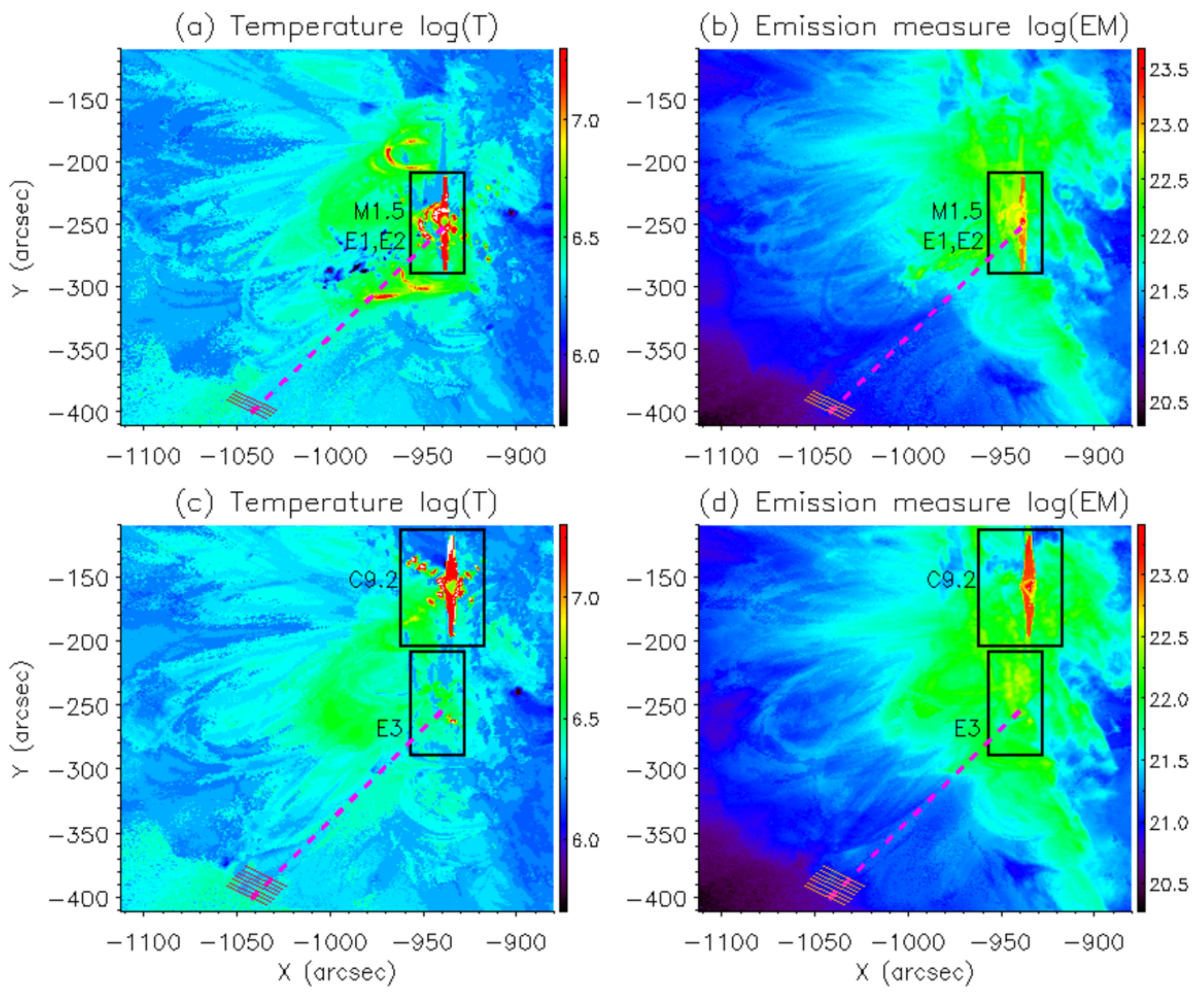}
\caption{Temperature and Emission Measure maps, in logarithmic scale, of the active region AR 1967. (a -- b) show $\log(T)$ and $\log(EM)$ maps during the eruptions (E1, E2) and M1.5 flare, while (c -- d) display the $\log(T)$ and $\log(EM)$ during the eruption E3 and C9.2 flare. The black boxes enclose the regions where these events occurred. The magenta dashed line indicates the central axis of the loops and the red lines, perpendicular to this line, represents the slits where we saw the oscillations, S0 -- S3 (a -- b) and S0 -- S6 (c -- d). The colour bar on the right side represents $\log(T)$ (a, c) and $\log(EM)$ (b, d).  \label{fig:maps}}
\end{center}
\end{figure}

The density of the plasma can be estimated by using the automated temperature and differential emission measure (DEM) analysis developed by \cite{ABSM13}. Thus, we made temperature $T$ (K) and emission measure $EM$ (cm$^{-5}$) maps considering the time at which the flare's energy flux peaked, i.e, 04:09 and 05:29 UT for M1.5 and C9.2 flares, respectively. Figure \ref{fig:maps} shows the Temperature and Emission Measure maps in the logarithmic scale, $\log(T)$ and $\log(EM)$ respectively, for the active region AR 1967 during the time in which the flare's peak was registered. Upper panels (a -- b) show the active region during the M1.5 flare together with eruptions E1 and E2, while bottom panels (c -- d) include the C9.2 flare and eruption E3. The black boxes enclose the regions where the events occurred, which also include the footpoint of the loops. The magenta dashed lines indicate the central axis of the loops. We highlighted with red and orange lines (perpendicular to the magenta dashed line), the slits S0 -- S3 (in panels (a -- b)) and S0 -- S6 (in panels (c -- d)) where we detected the oscillations. \\ 

With the information obtained from the $T$ and $EM$ maps, we calculated the number density $n_e$ (cm$^{-3}$) and then approximated the event density $\rho_{ev}$ (kg m$^{-3}$). During the M1.5 flare peak, $n_e=1.14 \times 10^9$ cm$^{-3}$ and $\rho_{ev}=1.14 \times 10^{-12}$ kg m$^{-3}$. At a later time, the C9.2 flare occurred further away from the location of the eruption E3. The density values during the C9.2 flare were $n_e=1.38 \times 10^9$ cm$^{-3}$ and $\rho_{ev}=1.38 \times 10^{-12}$ kg m$^{-3}$, while for the plasma in eruption E3 $n_e=2.06 \times 10^8$ cm$^{-3}$ and $\rho_{ev}=2.07 \times 10^{-13}$ kg m$^{-3}$. The number density values for the regions studied are presented in column 3 of Table \ref{t-ener}. In the same manner, we listed the eruptions and flares in column 1, indicating the wavelength in which they were observed, and their speeds in column 2. With these values and using the Equation (\ref{E-enev}), we found that energy density for the region during the M1.5 flare was in the range of $\sim 10^{-3} - 10^{-2}$ J m$^{-3}$. The energy density released by the C9.2 flare was $2.12 \times 10^{-1}$ J m$^{-3}$ and the eruption E3 was in the range of $\sim 10^{-6} - 10^{-4}$ J m$^{-3}$ (see column 4 of Table \ref{t-ener}). Although it is beyond the scope of the current study, the method reported by \cite{HK12} or \cite{CBST15} for determining DEM should be used to verify our results. \\

We estimated the kinetic energy density contained in the plasma at the locations of the slits (S0 -- S6), where we measured the oscillations \cite{GVSV13} 

\begin{equation}
	E_{os}=\frac{1}{4}\rho_{os} \omega^2 A_0^2,
	\label{E-enos}
\end{equation}

\noindent where $\rho_{os}$ is the density found in the slit from the $T$ and $EM$ maps, $A_0$ is the displacement amplitude, $\omega=2\pi/P$ is the angular frequency, and $P$ is the period. We listed $A_0$ and $P$ in Table \ref{t-fit} (columns 3 and 5). The corresponding number density for the slits was $\sim 10^7$ cm$^{-3}$ (see column 6 of Table \ref{t-ener}) and $\rho_{os} \sim 10^{-14}$ kg m$^{-3}$, during both M1.5 and C9.2 flares. Thus, the oscillations energy density during the M1.5 flare was $\sim 10^{-7} - 10^{-6}$ J m$^{-3}$. At the time of the C9.2 flare, this energy density was in the interval of $\sim 10^{-8} - 10^{-6}$ J m$^{-3}$ (see column 7 of Table \ref{t-ener}).

\begin{table}
\begin{center}
\caption{Number density and energy density estimated for the events and the oscillations.}
\label{t-ener}
\begin{tabular}{ccccccc} \hline \hline
Event & $v$ & $n_e$ & $E_{ev}$ & Slit & $n_e$ & $E_{os}$ \\
 & (km s$^{-1}$) & (cm$^{-3}$) & (J m$^{-3}$) & & (cm$^{-3}$) & (J m$^{-3}$) \\
(1) & (2) & (3) & (4) & (5) & (6) & (7) \\ \hline
	E1, E2  &  50.32 & 1.14$\times$10$^9$ & 1.44$\times$10$^{-3}$ & S0 & 4.18$\times$10$^7$ & 4.95$\times$10$^{-6}$ \\
	(171 \r{A}) & . & & & & & 1.72$\times$10$^{-6}$ \\ \hline
	E1, E2  & 222.18 & & 2.82$\times$10$^{-2}$ 
	& S1 & 4.93$\times$10$^7$ & 6.44$\times$10$^{-7}$ \\
	(171 \r{A}) & . & & & & & 8.62$\times$10$^{-7}$ \\ \hline
	E1, E2  & 102.92 & & 6.05$\times$10$^{-3}$
	& S2 & 3.88$\times$10$^7$ & 3.57$\times$10$^{-7}$ \\
	(131 \r{A}) & . & & & & & 4.45$\times$10$^{-7}$ \\ \hline
	E1, E2  & 195.87 & & 2.19$\times$10$^{-3}$
	& S3 & 4.61$\times$10$^7$ & 6.24$\times$10$^{-7}$ \\ 
	(304 \r{A}) & . & & & & & \\ \hline
	M1.5 flare & 216.43 & & 2.67$\times$10$^{-2}$ & & & \\ \hline
	\hline
	E3 (171 \r{A}) & 16.70 & 2.06$\times$10$^{8}$ & 2.89$\times$10$^{-5}$ & S0 & 3.62$\times$10$^7$ & 1.92$\times$10$^{-6}$ \\
	E3 (171 \r{A}) & 24.10 & & 6.02$\times$10$^{-5}$ & S1 & 4.32$\times$10$^7$ & 1.39$\times$10$^{-6}$ \\
	E3 (171 \r{A}) &  6.40 & & 4.24$\times$10$^{-6}$ & S2 & 4.26$\times$10$^7$ & 1.50$\times$10$^{-7}$ \\
	E3 (304 \r{A}) & 37.20 & & 1.43$\times$10$^{-4}$ & S3 & 4.15$\times$10$^7$ & 1.75$\times$10$^{-8}$ \\
	E3 (304 \r{A}) & 10.70 & & 1.18$\times$10$^{-5}$ & S4 & 4.34$\times$10$^7$ & 8.17$\times$10$^{-8}$ \\
	E3 (131 \r{A}) & 12.10 & & 1.52$\times$10$^{-5}$ & S5 & 4.11$\times$10$^7$ & 4.12$\times$10$^{-8}$ \\
	C9.2 flare & 553.03 & 1.38$\times$10$^9$ & 2.12$\times$10$^{-1}$ & S6 & 4.23$\times$10$^7$ & 5.17$\times$10$^{-8}$ \\ \hline
\end{tabular}
\end{center}
\end{table}

In the first part of the oscillatory motion (from 04:00 to 05:00 UT), the eruptions E1, E2, and the M1.5 flare occurred near the footpoints of the loops . In this case, the disturbance was likely focused on the central axis of the loops (see Figure \ref{fig:GOES}). The later part (between 05:00 and 06:30 UT) includes the eruption E3 and the C9.2 flare,  (see Figure \ref{fig:slitc92}). During this period, the oscillations were seen on the right side of the central axis of the loops. \\

Comparing the energy densities (columns 4 and 7 of Table \ref{t-ener}), we found that during the first period of our analysis, $E_{ev}$ was three to five orders of magnitude higher than $E_{os}$, which indicates that the energy contained in the events was enough to excite the perturbations in the slits S0 -- S3. As these events also occurred close to the central axis of the loops and have the same energy density; the eruptions E1, E2 and the M1.5 flare could probably contribute to producing the perturbations. At the later time, the eruption E3 happened close to the central axis of the loops and its energy was around $\sim 10^{-6} - 10^{-4}$ J m$^{-3}$. The energy released by the C9.2 flare was three to five orders greater than E3 ($\sim 10^{-1}$ J m$^{-3}$), despite being located $\approx 108.9$ Mm away. The eruption E3 stored energy just two orders of magnitude greater than the oscillations. However, the C9.2 flare released energy five to seven orders of magnitude greater than $E_{os}$, which suggest that, despite the long-distance, the C9.2 flare was the exciter agent of the perturbations on the slits S0 -- S6, between 05:00 and 06:30 UT.

\section{Conclusions} \label{sec:conclu}

In this paper, we report transverse oscillations in a loop structure in the active region AR 1967 during 04:00 -- 06:30 UT on January 28 2014. We believe that these oscillations were excited by multiple sources. The first flare (M1.5) which had the highest soft X-ray flux reported by the GOES instrument (see Figure \ref{fig:GOES}(d)) and the eruptions E1 and E2 (see movgr1967.m4v), excited oscillations with periodicities 4 -- 5 minutes that were clearly visible in the lower part (slits S0 -- S2) of the loop structure. The second flare (C9.2) soon after its initiations at about 05:25 UT reached its peak activity at 05:30 UT. The existing oscillation in the lower part of the loop  structure slits (S0 -- S2) not only continue to be seen after 05:25 UT but some oscillations with similar periodicities of 4 -- 6 minutes, became visible in the higher part of the loop structure (S3 -- S6). Despite the fact that C9.2 flare was located far ($\sim$ 109) Mm away from the loops, a fast wave emanating from the flaring site could have reached the loop-tops with a speed of about 553.03 km s$^{-1}$.  Thereafter, oscillations seem to be growing in amplitude with peak amplitude at $\sim$05:55 UT before suddenly dying out. Although around 05:45 UT a small increase in the X-ray flux was observed by the GOES instrument, this increase in the flux is relatively low compared to the other two flares. Thus, the oscillation is very likely influenced by the eruption E3 in addition to the flare C9.2 itself.\\

Another possible scenario for excitation of oscillations could be speculated as follows. Despite different energetics and the location of the flares, the loop oscillations seem to be {\it almost continuous} during the 2.5 hours that we examined with periodicity between 4 -- 6 minutes. Although these oscillations are not seen very clearly in all slits at all times, this could be due to fading and brightening of the loops in an active region. The fact that the periodicity is similar, there is a possibility that the field lines in the immediate surroundings of these loops in the magnetic arcade where these loops were embedded, provoked these oscillations. This is plausible because all magnetic structures in an active region are coupled with each other to varying extent and once flaring activities occur, the field lines within the active region set up resonances of their own as the waves from the flaring site get bounced around due to scattering, dissipation etc. If the resonance frequencies of the arcade and the loops are similar, it is possible to excite oscillations (see for details, \cite{HJ21}).\\

We find clear evidence of multiple oscillating threads (see Figure \ref{fig:FFTgv1}). A recent theoretical study by \cite{HJ21} suggests that coronal loops are part of a large flux system of an active region. It is possible that when there is an enormous disturbance such as a flare, the entire flux system gets impacted, but we only see oscillations in the bright loops. However, sometimes the nearby loops show similar oscillations when they lit-up. Thus, the oscillations of the most bright loop or loop structure are the response of a system of coupled loops and {\it oscillation of one loop can cause sympathetic vibration of the neighbouring loops} (see Figure \ref{fig:GOES}(d)) (see for details, \cite{HJ21}). Many previous observations have also implied that magnetic fields near a bright loop are often coupled (see for example, \cite{SB00,VNOD04,VAV09,JMH15}).\\

To understand the connection of the loop structures, their background and the flaring sites, we computed the FFT power in time domain at each pixel in the region of interest and displayed the dominant power within specific frequency bands. We found that for the time series of 2.5 hours (04:00 -- 06:30 UT), the dominant power is not only confined to the pixels located within the loop structure but it is distributed in other areas of the active region. This suggests that if flares are responsible for the oscillations, the flares excited the loop structure and the surrounding magnetic field regions with this periodicity or that the background coronal plasma has a broad band driver and the flares enhanced the power at specific frequency bands. However, these enhancements were not adequate to cause visible large amplitude oscillations in these areas.\\

We also see a close connection between eruptions during the time period and the initiation of oscillations. It is very likely that the flaring activity of M1.5 flare and the eruptions E1 and E2 associated with it were responsible for triggering the first few oscillatory motions in the lower slits (S0 -- S2) and the C9.2 flare perturbed the loops slightly higher and the perturbation must have arrived through fast waves since C9.2 flare was quite remote in relation to the loops location. However, this does not explain the peak in the oscillations between 05:45 -- 06:00 UT. Eruption E3 must have played a role in this peak. Despite the possibility of different excitation sources in the vicinity of the loops and the changing nature of oscillations, the periodicity of multiple oscillations is found to be between 4 -- 6 minutes.

\section*{acknowledgments}
	S.M.C.C. acknowledges financial support from the São Paulo Research Foundation FAPESP (Grants No. 2017/23026-6 and 2018/25306-9). R.J. and S.M.C.C. also acknowledge finances from MSRC (University of Sheffield, UK). V.J.-P. thanks FAPESP (Grant no. 2013/10559-5) for support.  We are also grateful to NASA/SDO, the AIA, the HMI, and GOES science teams.\\

\bibliographystyle{unsrt}
\bibliography{main.bib}

\end{document}